\documentclass[twocolumn,pra,showpacs]{revtex4-1}
\usepackage{amssymb}
\usepackage{amsmath}
\usepackage{graphicx}
\usepackage{subfigure}
\usepackage{natbib}
\usepackage{epsfig}
\usepackage{amsfonts}
\usepackage{mathrsfs}
\usepackage{CJK}
\usepackage[toc,page,title,titletoc,header]{appendix}
\usepackage{tikz}

\begin{document}

\title{The quasi-dark state and quantum interference in
  Jaynes-Cummings model with a common bath}

\author{Zhihai Wang}

\author{D.~L. Zhou}
\email{zhoudl72@aphy.iphy.ac.cn}

\address{Beijing National Laboratory for condensed matter physics,
  Institute of Physics, Chinese Academy of Sciences, Beijing, 100190,
  China}

\begin{abstract}
  Within the capacity of current experiments, we design a composite
  atom-cavity system with a common bath, in which the decay channels
  of the atom and the cavity mode interfere with each other. When the
  direct atom-cavity coupling is absent, the system can be trapped in
  a quasi-dark state (the coherent superposition of excited states for
  the atom and the cavity mode) without decay even in the presence of
  the bath. When the atom directly couples with the cavity, the
  largest decay rate of the composite system will surpass the sum of
  the two subsystems while the smallest decay rate may achieve $0$.
  This is manifested in the transmission spectrum, where the
  vacuum Rabi splitting shows an obvious asymmetric character.
\end{abstract}

\maketitle

\section{Introduction}
A quantum system in nature can not be absolutely isolated from its
surrounding environment~\cite{hp}. In addition, any quantum
measurement must introduce some interactions between the system and
the measurement instruments~\cite{vb1}. Therefore, to precisely
control and manipulate the quantum open systems is central task in quantum
information processes~\cite{hp,gmhuang,ma,vr}, such as quantum state
transfer and storage.

It is reported recently that, an efficient and long-lived quantum
memory was realized in a ring cavity proposal~\cite{jwpan}. One
basic element underlying the experimental scheme is that, the photons
can interact with the atoms more strongly with the assistance of the
ring cavity compared to the case in the free space. In this paper, we
theoretically propose a scheme as shown in Fig.~\ref{loop}, in which a
two-level atom is located in a dissipative cavity, the leaky light
from the cavity is reflected back to interact with the atom by four
high reflective mirrors (RMs). Therefore, the atom and the cavity mode
share a common bath which is composed of the light modes in the ring
cavity formed by the high RMs.  With the assistance of the common
bath, the interference between the decay channels for the atom and the
cavity mode leads to exotic behaviors which are great different from
the system of a singlet~\cite{aj,uw,yuting} or a
dimer~\cite{braun,majian,li,liao,swli} with two coupled or uncoupled
subsystems decaying independently.

To investigate the role of the interference effect, we reformulate the
traditional master equation, which is widely applied to study the
non-equilibrium dynamics of the open
system~\cite{th-1,as,lsb,ms,mj,bm,gg} under the secular
approximation~\cite{ct}. On one hand, when the direct atom-cavity
coupling is absent, the system will stay in a quasi-dark state which
is named by the analogy of dark state in the electromagnetic induced
transparency (EIT) phenomenon~\cite{EIT1}. For the quasi-dark state,
the decay processes of the atom and the cavity mode just cancel each
other due to the destructive interference, and it does not involve the
bare ground state. The existence of the quasi-dark state prevents the
system to achieve the thermal state equilibrium with the bath. On the
other hand, the Jaynes-Cummings (JC) type interaction between the atom
and the cavity mode~\cite{JC} gives birth to the entangled dressed
states. The constructive interference enhances the decay rate of the
symmetry dressed state which may achieve twice of the decay rate of
the atom or the cavity mode, while the destructive interference
suppresses the decay rate of the antisymmetry dressed state which may
even drop to $0$. A direct consequence is that, the vacuum Rabi
splitting~\cite{rj,gs,yifu,ss,th,fn,ty,ht} in the transmission
spectrum shows an obvious asymmetric character when the system is
driven by a probe light at zero temperature. The same asymmetric
splitting also occurs when the two level atom is replaced by the
single mode oscillator. However, it will behave differently at high
temperatures because of the different high excitation spectrum between
the atom-cavity system and the coupled oscillators system~\cite{FG}.

The paper is organized as follows. In Sec.~\ref{model}, we set up our
model and reformulate the traditional master equation to include the
interference terms between the decay of the cavity mode and the
spontaneous emission of the two level atom. We also show that the
interference slows down the decay of the system. In Sec.~\ref{qds}, we
demonstrate that the steady state of the atom-cavity system is the
quasi-dark state in the absence of direct atom-cavity interaction,
instead of the thermal equilibrium state. In Sec.~\ref{vrs}, we study
the vacuum Rabi splitting which shows an obvious asymmetry character
arising from the quantum interference and compare our model to the
coupled oscillators. The conclusions are drawn in
Sec.~\ref{conlusion}.

\section{Model and the master equation}
\label{model}
\begin{figure}[tbp]
\begin{centering}
\includegraphics[width=6 cm]{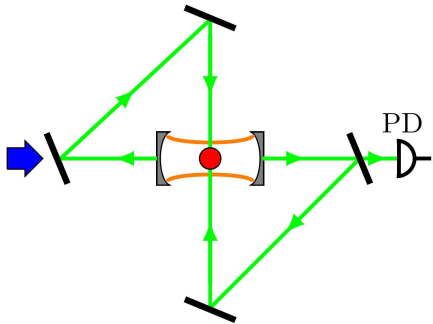}
\par \end{centering}
\caption{(Color online) The proposed experimental scheme.  A
    two-level atom (red dot) is located in a leaky cavity which
    supports a single mode electromagnetic field. The photons outside
    the cavity can be reflected by the mirrors and serve as a common
    bath shared by the atom and the cavity. The photon detector (PD) is
    applied to detect the number of photons in the cavity.}
\label{loop}
\end{figure}

We propose an experimental scheme as shown in Fig.~\ref{loop}. A
two-level atom is located in a high finesse cavity which supports a
single mode electromagnetic field. The light modes in the ring cavity
formed by four RMs construct the common bath shared by the cavity mode
and the two-level atom.

The Hamiltonian of the global system can be written as the sum of
three terms: $H=H_{JC}+H_{B}+H_{I}$, where
\begin{subequations}
  \begin{eqnarray}
    H_{JC} & = & \omega_{c} a^{\dagger} a + \frac{\omega_{0}}{2} \sigma_{z}
    + \lambda(a^{\dagger} \sigma^{-} + a \sigma^{+}), \label{eq:jc}\\
    H_{B} & = & \sum_{i} \omega_{i} b_{i}^{\dagger} b_{i}, \label{eq:bath}\\
    H_{I} & = & \sum_{i} \kappa_{i} (a b_{i}^{\dagger} + b_{i}
    a^{\dagger}) + \sum_{i} \xi_{i}(\sigma^{-} b_{i}^{\dagger} + b_{i}
    \sigma^{+}). \label{eq:int}
  \end{eqnarray}
\end{subequations}

The first term $H_{JC}$ is the Hamiltonian of our system, the JCM,
which describes a two-level atom interacting with the single mode
cavity photon under the rotating wave approximation. In
Eq.~(\ref{eq:jc}), $\omega_{c}$ is the frequency of the cavity mode,
$a$ is the annihilation operator of the mode. The two energy levels of
the atom are denoted as $|g\rangle$ and $|e\rangle$, and $\omega_{0}$
is the energy difference. The Pauli operators are defined as
$\sigma_{z}\equiv|e\rangle\langle{e}|-|g\rangle\langle{g}|$,
$\sigma^{-}\equiv|g\rangle\langle{e}|$, and
$\sigma^{+}\equiv|e\rangle\langle{g}|$. $\lambda$ is the coupling
strength between the atom and the cavity mode.

The second term $H_{B}$ describes the free terms of the photons in the
ring cavity, which act as a bath in our scheme. In
Eq.~(\ref{eq:bath}), $\omega_{i}$ is the frequency of the $i$-th mode
in the ring cavity and $b_{i}$ is its annihilation operator.

The third term $H_{I}$ describes the interactions between the system
and the bath. In Eq.~(\ref{eq:int}), $\kappa_{i}$ ($\xi_{i}$) is the
coupling strength between the atom (the cavity photons) and the $i$-th
mode of the bath.

Now, we study how the photon in the cavity mode decays into the bath.
Notice that there are two decay channels for the cavity photons. 
The cavity photons can either directly decay into the bath
or be absorbed by the atom and decay into the bath through the atomic
spontaneous emission . An intuitive idea is to sum the effects of the two
channels~\cite{fn} and the master equation can be formally written as
\begin{equation}
  \dot{\rho} = -i [H,\rho] + J_{1}(\omega_{c}) \mathcal{L}[a] + J_{2}(\omega_{0})
  \mathcal{L}[\sigma^{-}],  \label{eq:traditional}
\end{equation}
where $\mathcal{L}[Q]=(2Q\rho Q^{\dagger}-Q^{\dagger}Q\rho-\rho
Q^{\dagger}Q)$, and the spectrum functions $J_{1}(\omega)$ and
$J_{2}(\omega)$ are defined as ~\cite{aj,hefeng}
\begin{subequations}
  \begin{eqnarray}
    J_{1}(\omega) & = & \pi\sum_{j}\kappa_{j}^{2}\delta(\omega-\omega_{j}),
    \label{eq:j1} \\
    J_{2}(\omega) & = & \pi\sum_{j}\xi_{j}^{2}\delta(\omega-\omega_{j}).
    \label{eq:j2}
  \end{eqnarray}
  \label{j12}
\end{subequations}

However, the cavity mode and the atom share a common bath, and the
quantum interference between the two decay channels is completely
neglected in Eq.~(\ref{eq:traditional}).

To take into account the interference effect, we need to reformulate
the master equation. To this end, we first diagnose the Hamiltonian $H_{JC}$
for the the atom-cavity system. The ground state of $H_{JC}$ is a product state $|G\rangle=|0;g\rangle$ with eigen-energy
$E_{0}=-\omega_{c}/2$. In the resonance case
($\omega_{c}=\omega_{0}$), the energies for the excited states are
\begin{equation}
E_{n}^{\pm}=(n-1/2)\omega_{c}\pm \lambda\sqrt{n},
\label{JCL}
\end{equation}
and the corresponding
eigen-vectors are the dressed states
$|n,\pm\rangle=\left(\pm|n;g\rangle+|n-1;e\rangle\right)/\sqrt{2}$,
which are coherent superpositions of the product states $|n;g\rangle$
and $|n-1;e\rangle$.
Then the master equation can be derived
under the Markov and secular approximations with the standard steps. The detailed derivation is shown
in Ref.~\cite{ct}, the final result is obtained as

\begin{equation}
  \dot{\rho}_{cd}=-i(E_{c}-E_{d})\rho _{cd}+\sum_{k,l}\gamma ^{cdkl}\rho
  _{kl}  \label{eq:master}
\end{equation}
where $|c\rangle ,|d\rangle ,|k\rangle ,|l\rangle $ are the dressed
states of $H_{JC}$ with the eigen-energies $E_{c},E_{d},E_{k}$ and
$E_{l}$ respectively, $\rho_{cd}=\langle c|\rho | d\rangle$ and
$\rho_{kl}=\langle k|\rho | l\rangle$ are the elements of the reduced
density matrix for the atom-cavity system. In Eq. (\ref{eq:master}),
$\gamma^{cdkl}\equiv\sum_{i=1}^{4}\gamma _{i}$ with
\begin{widetext}
  \begin{subequations}
    \begin{eqnarray}
      \gamma _{1} &=&-\sum_{n}\left[
        J_{1}(\omega _{kn})\delta _{dl}a_{cn}^{\dagger }a_{nk}+J_{2}(\omega
        _{kn})\delta _{dl}\sigma _{cn}^{+}\sigma _{nk}^{-}
        +\sqrt{J_{1}(\omega _{kn})J_{2}(\omega _{kn})}\delta _{dl}\left(
          a_{cn}^{\dagger }\sigma _{nk}^{-}+\sigma _{cn}^{+}a_{nk}\right)
      \right]    \label{eq:ga1} \\
      \gamma _{2} &=&
      J_{1}(\omega _{kc})a_{ck}a_{ld}^{\dagger }+J_{2}(\omega _{kc})\sigma
      _{ck}^{-}\sigma _{ld}^{+}
      +\sqrt{J_{1}(\omega _{kc})J_{2}(\omega _{kc})}\left( a_{ck}\sigma
        _{ld}^{+}+\sigma _{ck}^{-}a_{ld}^{\dagger }\right)
      \label{eq:ga2} \\
      \gamma _{3} &=&-\sum_{n}\left[
        J_{1}(\omega _{ln})\delta _{ck}a_{ln}^{\dagger }a_{nd}+J_{2}(\omega
        _{ln})\delta _{ck}\sigma _{ln}^{+}\sigma _{nd}^{-}
        +\sqrt{J_{1}(\omega _{ln})J_{2}(\omega _{ln})}\delta _{ck}\left(
          a_{ln}^{\dagger }\sigma _{nd}^{-}+\sigma _{ln}^{+}a_{nd}\right)
      \right]    \\
      \gamma _{4} &=&
      J_{1}(\omega _{ld})a_{ck}a_{ld}^{\dagger }+J_{2}(\omega _{ld})\sigma
      _{ck}^{-}\sigma _{ld}^{+}
      +\sqrt{J_{1}(\omega _{ld})J_{2}(\omega _{ld})}\left( a_{ck}\sigma
        _{ld}^{+}+\sigma _{ck}^{-}a_{ld}^{\dagger }\right)
      \label{eq:ga4}
    \end{eqnarray}
    \label{eq:gamma}
  \end{subequations}
\end{widetext}
where $\omega _{ij}=E_{i}-E_{j}$ is the energy difference between
levels $i$ and $j$, and $A_{\alpha \beta} \equiv\langle\alpha
|A|\beta\rangle$ is the matrix element of operator $A$ in the dressed
state representation of the JCM. In the above equations, the terms
proportional to $\sqrt{J_{1}(\cdot)J_{2}(\cdot)}$ represent the
contribution from the quantum interference between the two decay
channels.

Under the Ohmic dissipation, the spectrum functions $J_{1}(\omega)$
and $J_{2}(\omega)$ of the bath are expressed as
\begin{subequations}
  \begin{eqnarray}
    J_{1}(\omega) & = & 2\pi \alpha_{1} \omega \exp(-\omega/\omega_{c1}),
    \\
    J_{2}(\omega) & = & 2\pi \alpha_{2} \omega \exp(-\omega/\omega_{c2}),
  \end{eqnarray}
\end{subequations}
where $\alpha_{1(2)}$ is the dissipation coefficients and
$\omega_{c1(c2)}$ denotes the cutoff frequencies.

\begin{figure}[tbp]
  \begin{centering}
    \includegraphics[width=8 cm]{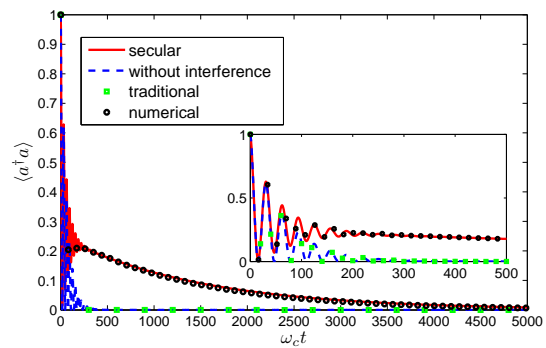}
    \par\end{centering}
  \caption{(Color online) The average photons number in the inner cavity
  as a function of the evolution time
  assuming the system is prepared in the state $|1;g\rangle\otimes|\bf{0}\rangle$  initially. The inset is
  the zoomed-in views. The parameters are set as
    $\omega_{0}=\omega_{c}=1, \lambda=0.1, \omega_{c1}=5,
    \omega_{c2}=8, \alpha_{1}=0.002, \alpha_{2}=0.001.$}
  \label{photon}
\end{figure}

In the end of this section, we point out the following two aspects.
Firstly, the size of the external cavity in our consideration is in
the order of $10cm$~\cite{jwpan}, while the inner cavity is in the
order of $nm$ to $\mu m$. Therefore, the external cavity is much
larger than the inner cavity, and can support a lot of electromagnetic
modes, which can be regarded as the environment. Although the photon
mode in the inner cavity and the photons modes in the environment are
all the bosonic modes, the coupling between the atom and photon mode
in the inner cavity is much stronger than that between the atom and
the environment, so we first diagnose the Hamiltonian of the system
($H_{JC}$) exactly, and regard the system-environment interaction as a
perturbation safely and apply the Markov approximation to discuss the
dynamics of the system. In Fig.~\ref{photon}, we
plot the average photons number in the inner cavity as a function of
the evolution time $t$ assuming the system is prepared in the state
$|1;g\rangle\otimes|\bf{0}\rangle$ initially, with $|\bf{0}\rangle$
representing that all of the bath modes are in their vacuum states.
For comparison, we also plot the results obtained by neglecting the
interference effect (that is omitting the terms proportional to
$\sqrt{J_1(\cdot)J_2(\cdot)}$ in Eqs.~(\ref{eq:gamma})) and the curve
obtained from the traditional master equation
(Eq.~(\ref{eq:traditional})). It is shown that, our results (secular)
coincide with the numerical results perfectly, which confirms the
effectiveness of Markov approximation. Besides, it is clearly
shown that the interference effect dramatically slows down the decay
of the whole system. The reason comes from the slow decaying of the
anti-symmetry dressed state $|1-\rangle$, whose decay behavior is
clearly shown in Sec.~\ref{vrs} and the appendix. Secondly, only the
bath modes which have the eigen frequencies around those of the lowest
dressed states couple to the system and the coupling strength is much
weaker than the atom-cavity coupling, i.e. $\lambda \gg
\alpha_{1},\alpha_{2}$. As a result, the system will maintain enough
coherence for a long time. As shown in Fig.~\ref{photon}, it exhibites
an obvious oscillation for the evolution time $\omega_c t< 200$ under
our parameters.

\section{Quasi-dark state}
\label{qds}
In this section, we firstly turn off the
interaction between the two level atom and the cavity mode, that is
$\lambda=0$. Then, the master equation~(\ref{eq:master}) degenerates
into a simple expression

\begin{eqnarray}
  \dot{\rho} & = & -i[\omega_{c} a^{\dagger} a + \frac{\omega_{0}}{2} \sigma_{z},\rho]\nonumber \\
  &&+J_{1}[2a\rho a^{\dagger}-a^{\dagger}a\rho-\rho a^{\dagger}a]\nonumber \\&&+J_{2}[2\sigma^{-}\rho\sigma^{+}-\sigma^{+}\sigma^{-}\rho-\rho\sigma^{+}\sigma^{-}]\nonumber \\
  &  & +\sqrt{J_{1}J_{2}}[2a\rho\sigma^{+}-\sigma^{+}a\rho-\rho\sigma^{+}a]\nonumber\\
  &&+\sqrt{J_{1}J_{2}}[2\sigma^{-}\rho a^{\dagger}-a^{\dagger}\sigma^{-}\rho-\rho a^{\dagger}\sigma^{-}],
  \label{simple}
\end{eqnarray}
where we write $J_1(\omega_c)$ as $J_1$ and $J_2(\omega_0)$ as $J_2$
for simple.

\begin{figure}[t]
  \centering \subfigure[]{
    \includegraphics[width=4cm]{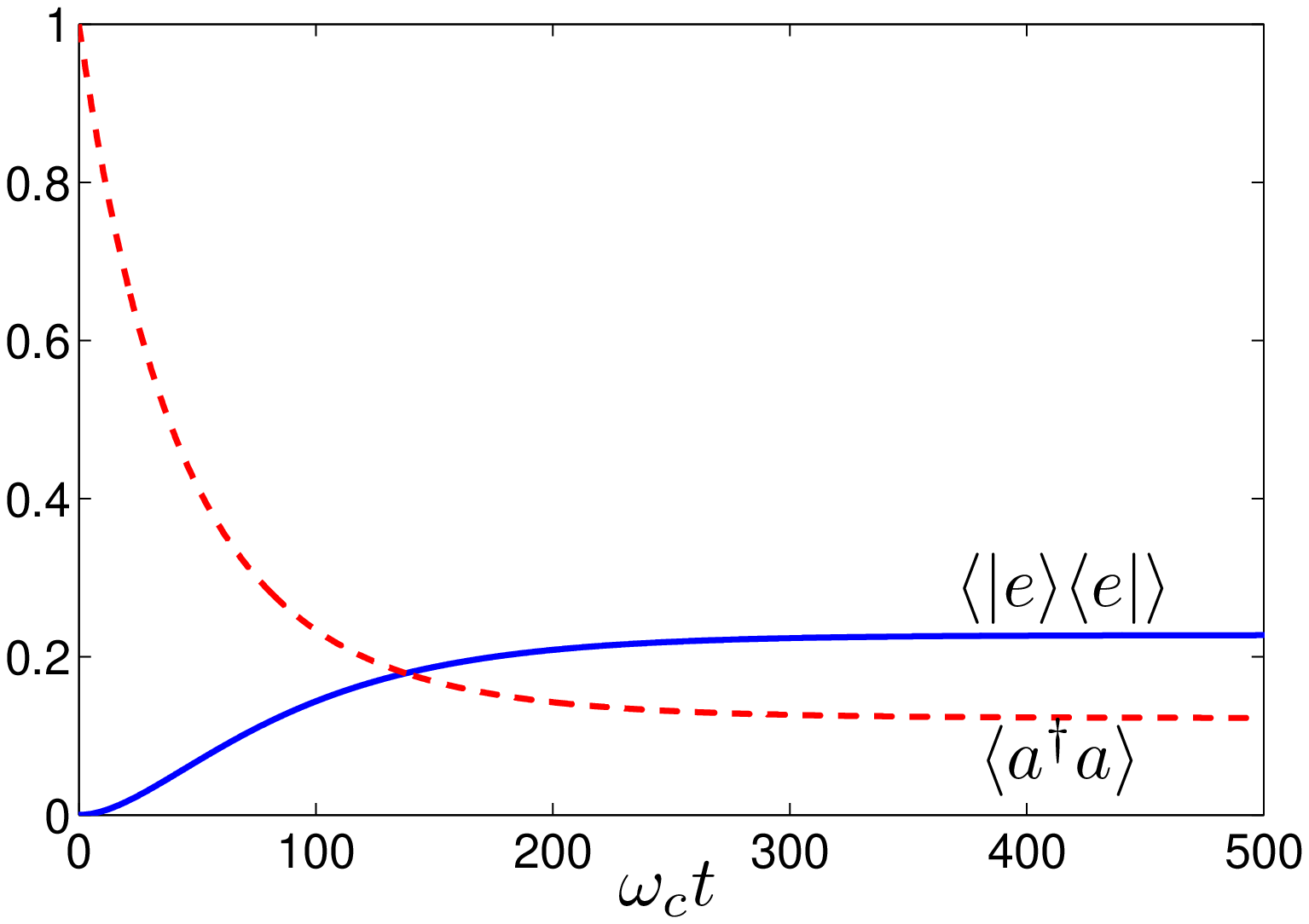}} \subfigure[]{
    \includegraphics[width=4cm]{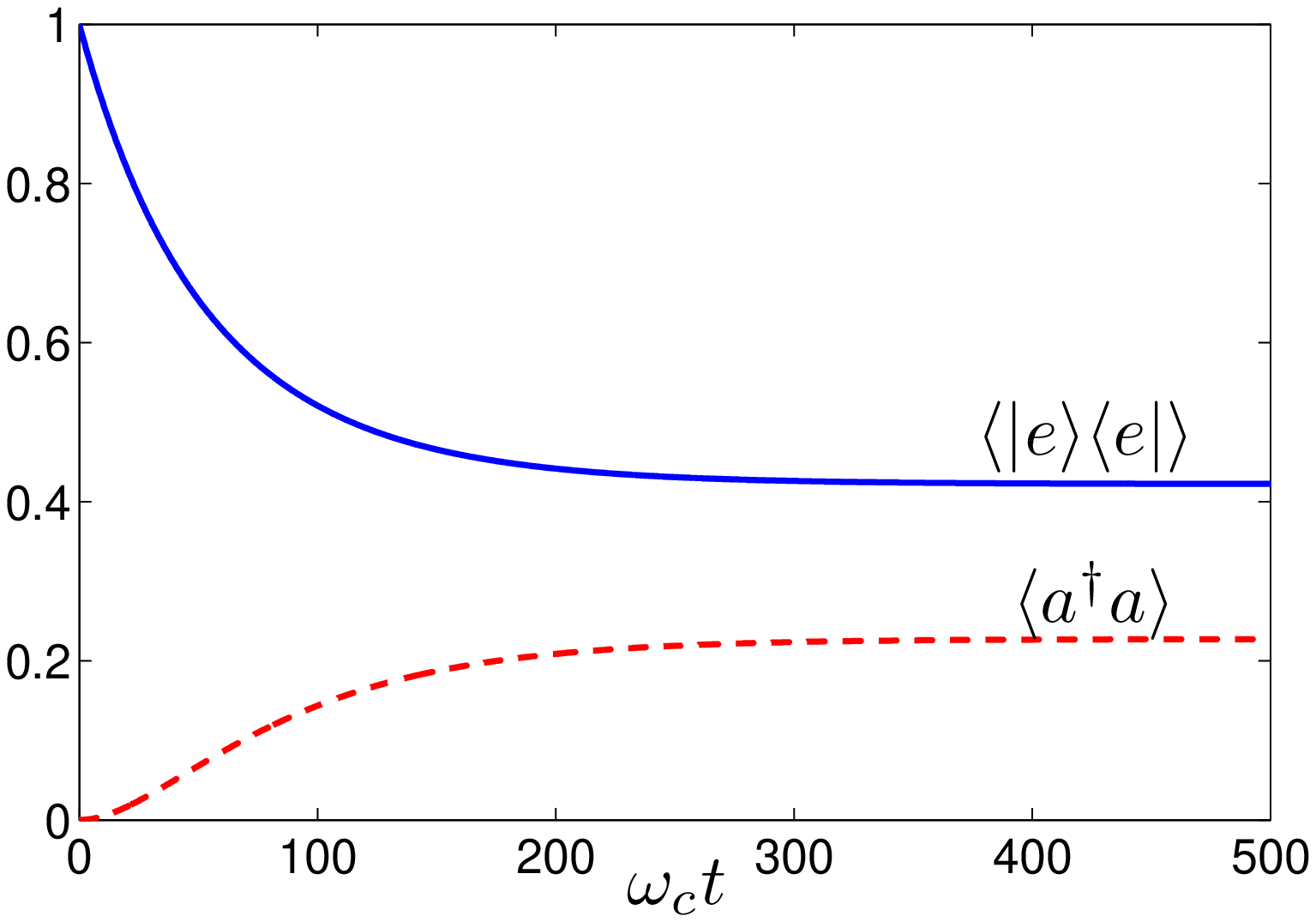}}
    \subfigure[]{\includegraphics[width=8cm]{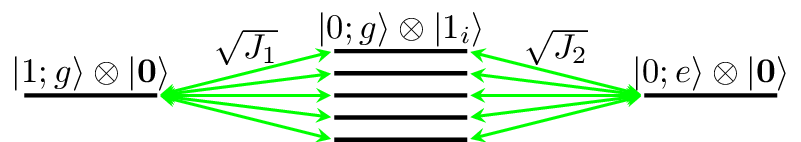}}
  \caption{(Color online) The average photons in the inner cavity and the
    probability for the atom in its excited state as a function of
    evolution time $t$ with the initial state being
    (a)$|1;g\rangle\otimes|\bf{0}\rangle$ and
    (b)$|0;e\rangle\otimes|\bf{0}\rangle$ respectively. (c) The sketch
    of the energy level transitions of the system.  The parameters are
    set as $\omega_{c}=\omega=1$, $\lambda=0$, $\omega_{c1}=5$,
    $\omega_{c2}=8$, $\alpha_{1}=0.002$, $\alpha_{2}=0.001$.}
  \label{steady}
\end{figure}

We prepare the system in the product state $|1;g\rangle \otimes |\bf{0}\rangle$ or
$|0;e\rangle\otimes|\bf{0}\rangle$ initially and investigate the dynamical
evolution of the system. Solving the above equation, we plot the curve of
the average photons in the inner cavity $\langle a^{\dagger }a\rangle$ and the
probability of the atom in its excited state
$\langle|e\rangle\langle{e}|\rangle$ as a function of the evolution
time in Fig.~\ref{steady}.

It is shown in the figure, the system will achieve a steady state
dependent of its initial state instead of the thermal state
equilibrium with the bath.  If the system is prepared in the state
$|1;g\rangle\otimes|\bf{0}\rangle$ initially, it satisfies
\begin{subequations}
  \begin{eqnarray}
    \langle a^{\dagger}a\rangle_s^{(g)}&=&\frac{J_2^{2}}{(J_1+J_2)^{2}}\\
    \langle |e\rangle \langle e|\rangle_s^{(g)}&=&\frac{J_1J_2}{(J_1+J_2)^{2}}
  \end{eqnarray}\label{1g1}
\end{subequations}
where $\langle A\rangle_s^{(g)}$ denotes the average value of the operator
$A$ over the steady state with the initial state being $|1;g\rangle\otimes|\bf{0}\rangle$.
On the contrary, if the system is prepared
in the state $|0;e\rangle\otimes|\bf{0}\rangle$ initially, it satisfies
\begin{subequations}
  \begin{eqnarray}
    \langle a^{\dagger}a\rangle_s^{(e)}&=&\frac{J_1J_2}{(J_1+J_2)^{2}}\\
    \langle |e\rangle \langle e|\rangle_s^{(e)}&=&\frac{J_1^{2}}{(J_1+J_2)^{2}}
  \end{eqnarray}\label{0e1}
\end{subequations}

These results can be understood from a simple physical picture as
shown in Fig.~\ref{steady}(c). The bath couples the two transition
arms $|1;g\rangle \otimes |{\bf 0}\rangle\leftrightarrow
|0;g\rangle\otimes|1_i\rangle$ and $|0;e\rangle\otimes |{\bf
  0}\rangle\leftrightarrow |0;g\rangle\otimes|1_i\rangle$
simultaneously, where $|1_i\rangle$ represents that the $ith$ mode in
the environment excites a photon, while the other modes are in their
vacuum states. Generally speaking, only the near resonant bath modes
contribute to the dynamics of the system, and the coupling intensities
can be regarded as constants for these bath modes in the Ohmic
spectrum situation. Furthermore, a discretization of Eq.~(\ref{j12})
gives that the coupling intensities of the two arms are proportional
to $\sqrt{J_1}$ and $\sqrt{J_2}$, respectively (as shown in
Fig.~\ref{steady}(c)). The destructive interference between the two
transitions occurs when the atom resonates with the cavity mode and
the atom-cavity system may be trapped in the excited states
\begin{equation}
  |D\rangle=\frac{\sqrt{J_{1}}|0;e\rangle-\sqrt{J_{2}}|1;g\rangle}{\sqrt{J_1+J_2}}
  \label{dark}
\end{equation}
without decay even in the presence of the bath. The state $|D\rangle$
is similar to the dark state in EIT phenomenon which is implemented
within the three-level or four-level systems, and we name it
quasi-dark state. A straight calculation gives
$\langle a^{\dagger}a\rangle_s^{(g)}=|\langle 1;g|D\rangle|^{2}|\langle D|1;g\rangle|^{2},
\langle |e\rangle \langle e|\rangle_s^{(g)}=|\langle 1;g| D\rangle|^{2}|\langle D|0;e\rangle|^{2},
\langle a^{\dagger}a\rangle_s^{(e)}=|\langle 0;e|D\rangle|^{2}|\langle D|1;g\rangle|^{2}$,
as well as
$ \langle |e\rangle \langle e|\rangle_s^{(e)}=|\langle 0;e|D\rangle|^{2}|\langle D|0;e\rangle|^{2}$. Therefore, we conclude that the system will stay in the quasi-dark state with certain probability dependent of the initial state when it achieves steady state. In other words, the quasi-dark state prevents the system to reach the thermal state equilibrium with the bath.

Actually, the dark state of the system can be obtained in a more intuitive way. To this end, we rewrite the master equation Eq.(\ref{simple}) in a more simple expression:
\begin{equation}
  \dot{\rho} =  -i[H_{0},\rho]+\mathcal{L}[P]
  \label{simp1}
\end{equation}
where
\begin{equation}
H_{0}=\omega_{c} a^{\dagger} a + \frac{\omega_{0}}{2} \sigma_{z},
\end{equation}
and
\begin{equation}
P=\sqrt{J_1}a+\sqrt{J_2}\sigma_{-}.
\end{equation}

Since the dark state does not decay even in the dissipative system, it
should be an eigenstate of $P$, we can directly write the dark state
$|D\rangle$ as shown in Eq.(\ref{dark}), which is not only the eigen
state of $P$ with a zero eigen value but satisfies that the
corresponding density matrix commutes with the Hamiltonian $H_{0}$,
and then $\dot{\rho}=0$. Physically speaking, it is the interference
effect between the different dissipation channels that leads to the
existence of the dark state. Arising from the same mechanism, the high
fidelity dark entangled steady states can also be rapidly generated in
interacting Rydberg atoms system~\cite{DD}.

\section{Vacuum Rabi splitting}
\label{vrs}
To further explore the effect of a common bath, we study the
transmission spectrum of the system considering the direct atom-cavity
coupling ($\lambda\neq0$). To this aim, we drive the cavity by an
external field and the action of the driven field on the cavity mode
is described by
\begin{equation}
  H_{driven}=\eta(a^{\dagger}e^{-i\omega_{d}t}+ae^{i\omega_{d}t})
\end{equation}
where $\eta$ denotes the intensity of the driven field and
$\omega_{d}$ its frequency.

The driven field induces the transition between the ground state and
the excited states, while the dissipation causes the excited states to
decay to the ground state.  When the evolution time is long enough,
the system may reach a steady state.  To investigate the behavior of
the steady state, we eliminate the time dependence from the
Hamiltonian through the unitary transformation
$U(t)=\exp[i(a^{\dagger}a+\sigma_{z}/2+\sum_{i}b_{i}^{\dagger}b_{i})\omega_{d}t]$.
The atom-cavity Hamiltonian in the rotating frame becomes
\begin{eqnarray} \label{eq:rtf} \mathcal{H} & = &
  i\frac{\partial}{\partial t}U(t)U^{\dagger}(t)+U(t)
  \left(H_{JC}+H_{driven}\right)U^{\dagger}(t)  \notag \\
  & = &
  \Delta_{1}a^{\dagger}a+\frac{\Delta_{2}}{2}\sigma_{z}+\lambda(a^{\dagger}
  \sigma^{-}+a\sigma^{+})+\eta(a^{\dagger}+a)  \notag \\
\end{eqnarray}
where $\Delta_{1(2)}=\omega_{c(0)}-\omega_{d}$ is the detuning between
the cavity (atom) and the driven field. The last term can be regarded
as a perturbation whenever $\eta\ll\omega_{c},\omega_{0},\lambda,$
i.e., weak driven field.  Then the master equation can be written as
\begin{equation}
  \frac{d}{dt}\rho_{cd} =  -i\langle c[\mathcal{H},\rho]|d\rangle+%
  \sum_{k,l}\gamma^{cdkl}\rho_{kl}.
\end{equation}

\begin{figure}[tbp]
  \begin{centering}
    \includegraphics[width=8 cm]{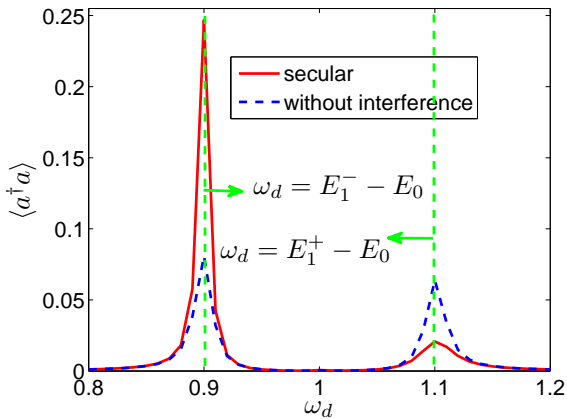}
    \par\end{centering}
  \caption{(Color online) The steady state of the system driven by an
    external field.  The parameters are set as
    $\omega_{0}=\omega_{c}=1, \lambda=0.1, \omega_{c1}=5,
    \omega_{c2}=8, \alpha_{1}=0.002, \alpha_{2}=0.001,\eta=0.005.$}
  \label{drive}
\end{figure}

Therefore, the average photon number over the steady state is obtained
numerically by finding the density matrix satisfying $d\rho_{cd}/dt=0$
for any eigen states $|c\rangle$ and $|d\rangle$ of $H_{JC}$. The results are shown in
Fig.~\ref{drive}, where the average photon number reaches its peaks
when the driven field is just resonant with the energy difference
between the dressed states $|1,\pm\rangle$ and the ground state
$|G\rangle$. This effect is the vacuum Rabi splitting.

In Fig.~\ref{drive}, we also plot the results given by neglecting the
contribution from quantum interference. It is shown that the double
peaks exhibit an obvious asymmetric character if the quantum
interference between two decay channels is taken into
consideration. As observed, the peak corresponding to
$\omega_{d}=E_{1}^{-}-E_{0}$ is elevated while the peak corresponding
to $\omega_{d} =E_{1}^{+}-E_{0}$ is suppressed compared with the case
that neglects the effect of quantum interference.

Now let us explain the physics underlying the exotic phenomenon and
analysis how the quantum interference effect affects the decay rate of
the dressed states in the following two steps.

Firstly, we neglect the interference between the two decay
channels. That is, the atom and the cavity experience dissipation
independently. As discussed above, the atom-cavity coupling dressed
them and the eigen energies of the dressed states $|1,\pm\rangle$ are
$\omega_{1,\pm}$. Following the master equation~(\ref{eq:master}) and
setting the terms proportional to $\sqrt{J_1(\cdot)J_2(\cdot)}$ to
zero in Eqs.~(\ref{eq:gamma}), the decay rate of the dressed states
$|1,\pm\rangle$ are both the sum of the decay rate of the atom and the
cavity mode, that is,
$\Gamma^{1,\pm}=J_{1}(\omega_{1\pm,0})+J_{2}(\omega_{1\pm,0})$. In the case of
Ohmic spectrum, it satisfied $\Gamma^{1,+}\approx\Gamma^{1,-}$, therefore, we
obtain two nearly symmetrical peaks in Fig.~\ref{drive}.
\begin{figure}[tbp]
  \centering \subfigure[]{
    \includegraphics[width=6cm]{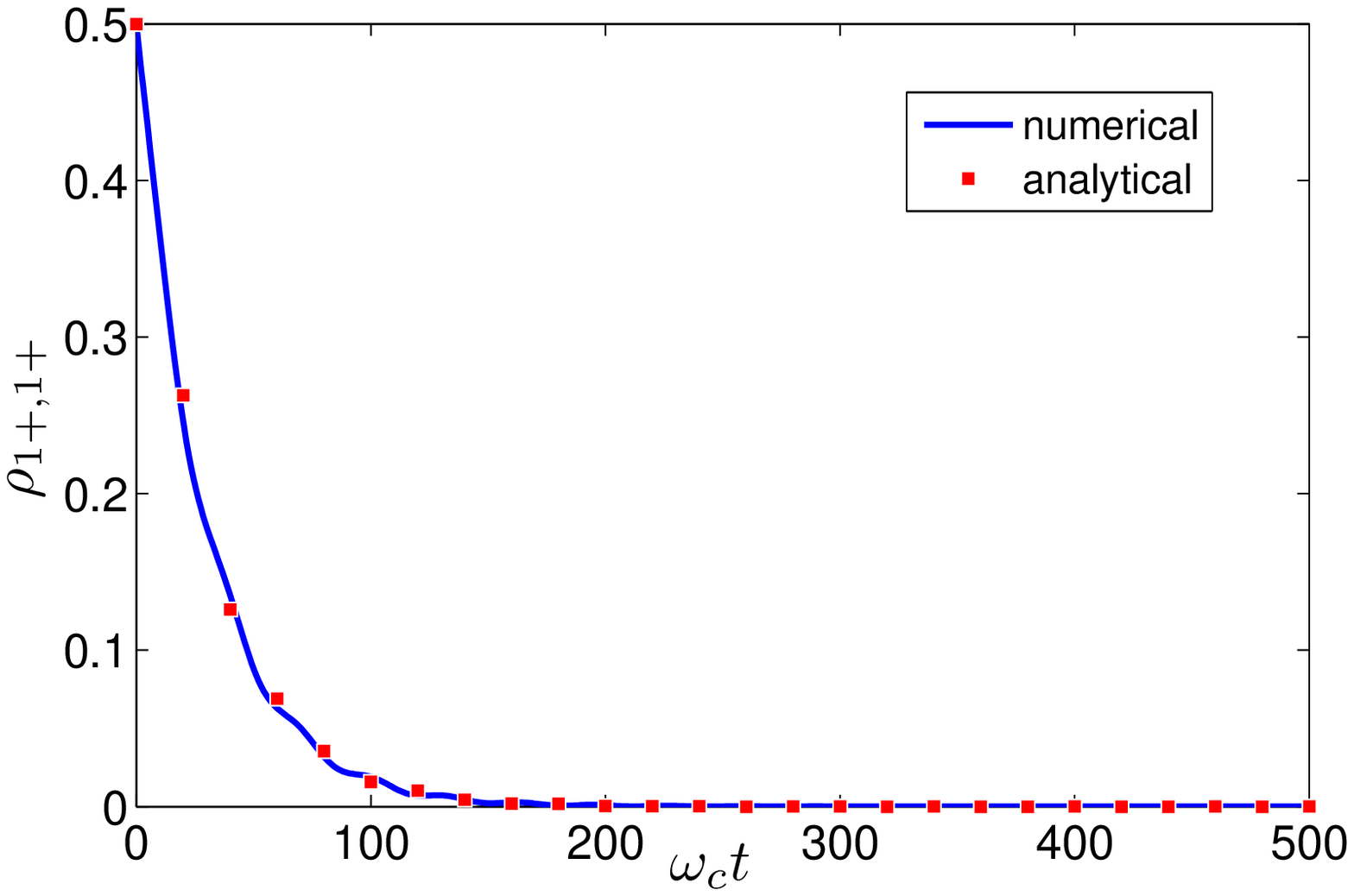}} \subfigure[]{
    \includegraphics[width=6cm]{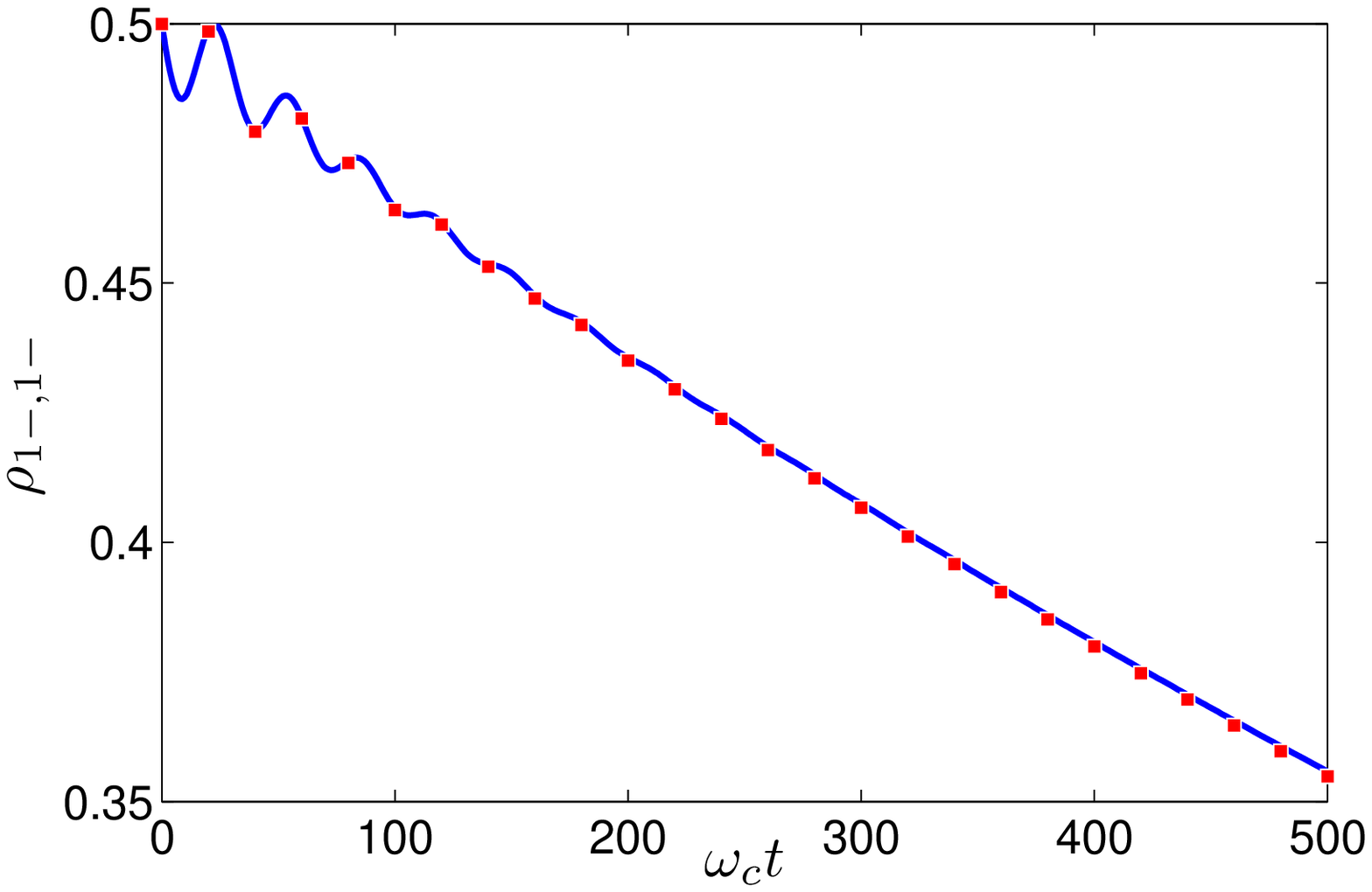}}
  \caption{(Color online) The time evolution of matrix elements $\rho_{1+,1+}$
and $\rho_{1-,1-}$. The solid line: the numerical solution, The solid square: the analytical
solution whose detail derivation is shown in the appendix. The parameters are set as
 $\omega_{0}=\omega_c=1$, $\lambda=0.1$, $\omega_{c1}=5,$ $\omega_{c2}=8$,
$\alpha_{1}=0.002,$ $\alpha_{2}=0.001.$ The system is initially prepared in the state $|1;g\rangle \otimes |\bf{0}\rangle$.}
  \label{dress}
\end{figure}

Then, we furthermore consider the quantum interference between the two
decay channels. Arising from the constructive interference, the decay
rate of the symmetric dressed state $|1,+\rangle$ becomes $\big(
\sqrt{J_{1}(\omega_{1+,0})}+\sqrt{J_{2}(\omega_{1+,0})} \big)^{2}$
which even surpasses the sum of the decay rate of subsystems. On
contrary, due to the destructive interference, the decay rate of the
antisymmetric dressed state $|1,-\rangle$ is $\big(
\sqrt{J_{1}(\omega_{1-,0})}-\sqrt{J_{2}(\omega_{1-,0})} \big)^{2}$
which may even achieve $0$ under some special parameters (for example
$J_{1}(\omega_{1-,0})=J_{2}(\omega_{1-,0})$). It implies that the
antisymmetry state will have an infinite lifetime.

The above analysis implies that the symmetric dressed state
$|1,+\rangle$ has a much larger decay rate than that of the antisymmetric
dressed state $|1,-\rangle$. The same conclusion is also shown in
Fig.~\ref{dress}, where we plot the probability for the system in the
symmetric and antisymmetric dressed states (the time evolution of the
density matrix elements $\rho_{1+,1+}$ and $\rho_{1-,1-}$), assuming
the system is prepared in the state $|1;g\rangle\otimes|\bf{0}\rangle$
initially. It is clearly that the symmetric state $|1,+\rangle$ decays
more faster than the antisymmetric state $|1,-\rangle$. Apart from the
numerical result, an analytical result based on iteration calculations
is shown in the appendix.

Now, the asymmetric vacuum Rabi splitting can be explained from the
different decay rates of the states $|1,+\rangle$ and
$|1,-\rangle$. Because the state $|1,-\rangle$ has a smaller decay rate,
the steady state will have a larger component. This will give a
stronger signal of the average photon number. In contrary, the state
$|1,+\rangle$ has a larger decay rate, so the steady state will have a
smaller component and it will give a weaker signal.

As is well known, the eigen states of the JC model consist a bare
ground state and pairs of dressed states, which are anti-symmetric and
symmetric states when the cavity mode is resonant with the atom. From
the above discussions, it can be concluded that any anti-symmetric
dressed state $|n,-\rangle$ for both $n=1$ and $n\neq1$ will have a
smaller decay rate than the corresponding symmetric dressed state
$|n,+\rangle$.

Actually, the same asymmetric Rabi splitting is also given when the
two level atom is replaced by a single mode harmonic oscillator. This
similarity can be clarified by investigating the low energy levels of
the system. To this end, we write the Hamiltonian of coupled resonant
oscillators as
\begin{eqnarray}
H_{co}&=&\omega a^{\dagger} a+b^{\dagger} b+\lambda(a^{\dagger} b+b^{\dagger}a)\nonumber\\
&=&(\omega+\lambda)A^{\dagger}A+(\omega-\lambda)B^{\dagger}B.
\end{eqnarray}
where $a$ and $b$ are the annihilation operators for the two coupled
oscillators respectively, and the new set of bosonic operators $A, B$
are defined as $A=(a+b)/\sqrt{2}$ and $B=(a-b)/\sqrt{2}$. Diagnosing
the Hamiltonian $H_{co}$ when the coupling strength is smaller than
the frequencies of the oscillators ($\lambda<\omega$), the eigen
energies can be obtained as
\begin{equation}
E_{m_1,m_2}=m_1(\omega+\lambda)+m_2(\omega-\lambda)
\label{coupleo}
\end{equation}
where $m_1$ and $m_2$ are integers. It is obvious that energies of the
first three energy levels are
$E_{0,0}=0,E_{0,1}=\omega-\lambda,E_{1,0}=\omega+\lambda$ and the
splitting is $E_{1,0}-E_{0,1}=2\lambda$ which is same as that of the
JC model. Therefore, we will obtain similar asymmetric peaks in the
transmission spectrum at zero temperature.

However, the higher energy levels of the coupled oscillators (as shown in Eq.~(\ref{coupleo})) 
are much different from those of the JC model which are shown in Eq.~(\ref{JCL}). Therefore, the
splitting will behave differently in the two systems in finite temperature.

\section{Conclusions and remarks}
\label{conlusion}
In this paper, we discuss the dissipation of the interacting
atom-cavity system when the atom and the cavity mode share a common
bath. We regard the spectrum of the environment as Ohmic spectrum
which is the major decoherence source often found in the qubit's
environment~\cite{aj,uw}. We can also choose other type environment,
but the quantum interference effect will not be changed
quantitatively. To investigate the effect of the common bath, we
reformulated the master equation and obtained simple expressions as
shown in Eqs.~(\ref{simple},\ref{simp1}). Actually, it can also be
written as the form of Fokker-Planck equation in coherence state
representation~\cite{carmichael}. Solving the Fokker-Planck equation,
we need not to truncate the photons number and it is convenient to
deal with the system with large Hilbert space. However, it includes no
more physics than the operator form of the master equation because
they are indeed the same equations in different representations.

In summary, we have proposed a scheme to realize the JCM with a common
bath within the current experimental capabilities. In our system, the
common bath induces a quasi-dark state which does not decay and
prevents the system to equilibrate with the bath. Besides, the decay
processes of the atom and the cavity interfere with each other and
the constructive interference leads the symmetric dressed
state decaying much faster than the antisymmetric dressed state.
As a result, the vacuum Rabi splitting in the transmission spectrum
shows an obvious asymmetric character. Furthermore, the robustness of
the antisymmetric dressed state may be applied in quantum information
process, such as the storage of the quantum state in quantum network.

\begin{acknowledgments}
  This work was supported by National NSF of China (Grant Nos.
  10975181 and 11175247) and NKBRSF of China (Grant No. 2012CB922104).
\end{acknowledgments}

\appendix%\appendixpage
%\addcontentsline{toc}{section}{Appendices}\markboth{APPENDICES}{}
\section{The iteration solution for $\rho_{1+,1+}$ and $\rho_{1-,1-}$ }
In the main text, we have solved the master equation numerically and
obtained the time evolution of the matrix elements $\rho_{1+,1+}$ and
$\rho_{1-,1-}$ as shown in Fig.~4. Now, we will give some analytical
results based on the iteration calculation. To this end, we restrict
our consideration in no excitation and only one excitation subspaces
which are spanned by the ground state $|G\rangle$ and the dressed
states $|1,\pm\rangle$ of the JCM. Then, Utilizing the master Eq.~(4)
in the main text, the equations for $\rho_{1\pm,1\pm}$ are written as
\begin{subequations}
\begin{eqnarray}
\dot\rho_{1+,1+} & = & \gamma^{1+,1+,1+,1+}\rho_{1+,1+}+\gamma^{1+,1+,1+,1-}\rho_{1+,1-}\notag\\ &&+\gamma^{1+,1+,1-,1+}\rho_{1-,1+},\label{1p1p}\\
\dot \rho_{1-,1-} & = & \gamma^{1-,1-,1-,1-}\rho_{1-,1-}+\gamma^{1-,1-,1+,1-}\rho_{1+,1-}\notag\\&&
+\gamma^{1-,1-,1-,1+}\rho_{1-,1+},\label{1n1n}\\
\dot\rho_{1+,1-} & = & -i(\omega_{1+}-\omega_{1-})\rho_{1+,1-}+\gamma^{1+,1-,1+,1-}\rho_{1+,1-}\notag \\&&+\gamma^{1+,1-,1+,1+}\rho_{1+,1+}+\gamma^{1+,1-,1-,1-}
\rho_{1-,1-},\notag \\\label{1p1n} \\
\dot\rho_{1-,1+} & = & -i(\omega_{1-}-\omega_{1+})\rho_{1-,1+}+\gamma^{1-,1+,1-,1+}\rho_{1-,1+}\notag \\&&+\gamma^{1-,1+,1+,1+}\rho_{1+,1+}
+\gamma^{1-,1+,1-,1-}\rho_{1-,1-}.\notag \\\label{1n1p}
\end{eqnarray}
\end{subequations}
Now, we solve the matrix elements $\rho_{1+,1+}$ and $\rho _{1-,1-}$ in two steps.
Firstly, we neglect the dependence on $\rho_{1+,1+}$ and $\rho _{1-,1-}$ of  $\rho_{1+,1-}$
and $\rho _{1-,1+}$, that is, the last two terms in Eq.~(\ref{1p1n}) and Eq.~(\ref{1n1p}). Then we obtain
\begin{subequations}
\begin{eqnarray}
\rho_{1+,1-} & = & \rho_{1+,1-}(0)e^{[\gamma^{1+,1-,1+,1-}-i(\omega_{1+}-\omega_{1-})]t}\\
\rho_{1-,1+} & = & \rho_{1-,1+}(0)e^{[\gamma^{1-,1+,1-,1+}+i(\omega_{1+}-\omega_{1-})]t}
\end{eqnarray}
\end{subequations}

Secondly, substituting the last two equations back into Eq.~(\ref{1p1p}) and  Eq.~(\ref{1n1n}), we finally obtain
the matrix elements $\rho_{1+,1+}$ and $\rho _{1-,1-}$ as
\begin{widetext}
\begin{subequations}
\begin{eqnarray}
\rho_{1+,1+} & = & \left[\begin{array}{c}
\gamma^{1+,1+,1+,1-}\rho_{1+,1-}(0)\frac{\exp\{[\gamma^{1+,1-,1+,1-}-\gamma^{1+,1+,1+,1+}-i(\omega_{1+}-\omega_{1-})]t\}-1}{\gamma^{1+,1-,1+,1-}-\gamma^{1+,1+,1+,1+}-i(\omega_{1+}-\omega_{1-})}\\
+\gamma^{1+,1+,1-,1+}\rho_{1-,1+}(0)\frac{\exp\{[\gamma^{1-,1+,1-,1+}-\gamma^{1+,1+,1+,1+}+i(\omega_{1+}-\omega_{1-})]t\}-1}{\gamma^{1-,1+,1-,1+}-\gamma^{1+,1+,1+,1+}+i(\omega_{1+}-\omega_{1-})}\\
+\rho_{1+,1+}(0)
\end{array}\right]e^{\gamma_{1+,1+,1+,1+}t}\\
\rho_{1-,1-} & = & \left[\begin{array}{c}
\gamma^{1-,1-,1+,1-}\rho_{1+,1-}(0)\frac{\exp\{[\gamma^{1+,1-,1+,1-}-\gamma^{1-,1-,1-,1-}-i(\omega_{1+}-\omega_{1-})]t\}-1}{\gamma^{1+,1-,1+,1-}-\gamma^{1-,1-,1-,1-}-i(\omega_{1+}-\omega_{1-})}\\
+\gamma^{1-,1-,1-,1+}\rho_{1-,1+}(0)\frac{\exp\{[\gamma^{1-,1+,1-,1+}-\gamma^{1-,1-,1-,1-}+i(\omega_{1+}-\omega_{1-})]t\}-1}{\gamma^{1-,1+,1-,1+}-\gamma^{1-,1-,1-,1-}+i(\omega_{1+}-\omega_{1-})}\\
+\rho_{1-,1-}(0)
\end{array}\right]e^{\gamma_{1-,1-,1-,1-}t}.
\end{eqnarray}
\end{subequations}
\end{widetext}
where we have defined $\gamma s$ in the main text. In Fig.~\ref{dress} ,we find that the results from the iteration calculation agree with numerical results very well.


\begin{thebibliography}{99}
\bibitem{hp} H. P. Breuer and F. Petruccione, \textit{The Theory of
    Open Quantum Systems} (Oxford University Press, Oxford, London,
  2002).
\bibitem{vb1}V. Braginsky and F. Khalili, \textit{Quantum Measurement}
  (Cambridge University Press, London, 1992).

\bibitem{gmhuang} G. M. Huang, T. J. Tarn, and J. W. Clark,
  J. Math. Phys. \textbf{24}, 2608 (1983).

\bibitem{ma}M. A. Nielsen and I. L. Chuang, \textit{Quantum
    Computation and Quantum Information} (Cambridge University Press,
  Cambridge, U.K., 2000).

\bibitem{vr} V. Ramakrishna and H. Rabitz, Phys. Rev. A \textbf{54},
  1715 (1996).

\bibitem{jwpan} X. H. Bao, A. Reingruber, P. Dietrich, J. Rui,
  A. Duck, T. Strassel, L. Li, N. L. Liu, B. Zhao, and J. W. Pan,
  Nature Phys. \textbf{8}, 512 (2012)

\bibitem{aj} A. J. Leggett, S. Chakravarty, A. T. Dorsey,
  M. P. A. Fisher, A. Garg, and W. Zwerger,
  Rev. Mod. Phys. \textbf{59}, 1 (1987).

\bibitem{uw} U. Weiss, \textit{Quantum Dissipative Systems, third
    Edition}, (World Scientific Publishing, 2008).

\bibitem{yuting}T. Yu and J. H. Eberly, Phys. Rev. Lett. \textbf{93},
  140404 (2004).

\bibitem{braun}D. Braun, Phys. Rev. Lett. \textbf{89}, 277901 (2002).

\bibitem{majian}J. Ma, Z. Sun, X. G. Wang, and F. Nori, Phys. Rev. A
  \textbf{85}, 062323 (2012).

\bibitem{li}J. Li and G. S. Paraoanu, New J. Phys. {\bf 11},
113020 (2009).


\bibitem{liao}J. Q. Liao, J. F. Huang, and L. M. Kuang, Phys. Rev. A
  \textbf{83}, 052110 (2011).

\bibitem{swli}S. W. Li, L. P. Yang, and C. P. Sun, Arxiv.~1303, 1266v1
  (2013).

\bibitem{th-1} T. H$\ddot{\mathrm{a}}$yrynen, J. Oksanen, and
  J. Tulkki, Phys. Rev. A \textbf{83}, 013801 (2011).

\bibitem{as} A. Stokes, A. Kurcz, T. P. Spiller, and A. Beige,
  Phys. Rev. A \textbf{85}, 053805 (2012).

\bibitem{lsb} L. S. Bishop, E. Ginossar, and S. M. Girvin,
  Phys. Rev. Lett.  \textbf{105}, 100505 (2010).

\bibitem{ms} M. Scala, B. Militello, A. Messina, J. Piilo, and S.
  Maniscalco, Phys. Rev. A \textbf{75}, 013811 (2007).

\bibitem{mj} M. J. Bhaseen, J. Mayoh, B. D. Simons, and J. Keeling,
  Phys. Rev. A \textbf{85}, 013817 (2012).

\bibitem{bm} B. Masashi, J. Phys. A: Math. Theor. \textbf{43}, 335305
  (2010).

\bibitem{gg} G. Gangopadhyay, S. Basu, and D. S. Ray, Phys. Rev. A
  \textbf{47}, 1314 (1993).

\bibitem{ct} C. Cohen-Tannouji, J. Dupont-Roc, and G. Grynberg,
  \textit{Atom-Photon Interactions: Basic Process and
    Applications}(John,Wiley \& Sons,New York,1998).

\bibitem{EIT1} S. E. Harris, Phys. Today \textbf{50}, 36 (1997).

\bibitem{JC} E. T. Jaynes and F. W. Cummings, Proc IEEE \textbf{51},89
  (1963).



\bibitem{th} T. H$\ddot{\mathrm{u}}$mmer, G. M. Reuther,
  P. H$\ddot{\mathrm{a%
    }}$nggi, and D. Zueco, Phys. Rev. A \textbf{85}, 052320 (2012).

\bibitem{fn} F. Nissen, S. Schmidt, M. Biondi, G. Blatter,
  H. E. T$\ddot{%
    \mathrm{u}}$reci, and J. Keeling, Phys. Rev. Lett. \textbf{108},
  233603 (2012).

\bibitem{gs} G. S. Agarwal, Phys. Rev. Lett. \textbf{57}, 1732 (1984).

\bibitem{rj} R. J. Thompson, G. Rempe, and H. J. Kimble,
  Phys. Rev. Lett.  \textbf{68}, 1132 (1992).

\bibitem{yifu} Y. F. Zhu, D. J. Gauthier, S. E. Morin, Q. L. Wu, H. J.
  Carmichael, and T. W. Mossberg, Phys. Rev. Lett. \textbf{64}, 2499
  (1990).

\bibitem{ss} S. Savasta, R. Saija, A.Ridolfo, O. D. Stefano, P. Denti,
  and F. Borghese, ACS. Nano. \textbf{4}, 6369 (2010).

\bibitem{ty} T. Yoshie, A. Scherer, J. Hendrickson, G. Khitrova, H. M.
  Gibbs, G. Rupper, C. Ell, O. B. Shchekin, and D. G. Deppe, Nature,
  \textbf{432}, 200 (2004).

\bibitem{ht} H. Toida, T. Nakajima, and S. Komiyama, Phys. Rev. Lett.
  \textbf{110}, 066802 (2013).

 \bibitem{FG} F. Galve, G. L. Giorgi, and R. Zambrini, Phys. Rev. A
  \textbf{81}, 062117 (2010)

\bibitem{hefeng} H. F. Wang, S. Ashhab, and F. Nori, Phys. Rev. A
  \textbf{83}, 062317 (2011).

\bibitem{DD} D. D. Bhaktavatsala Rao and K. M{\o}lmer, Phys. Rev. Lett.
  \textbf{111}, 033606 (2013).

\bibitem{carmichael}  H. J. Carmichael, \textit{Statistical Methods in Quantum Optics I}
(Springer, 1999)




\end{thebibliography}
\end{document}